\documentclass[12pt]{article}
\usepackage{amsmath}
\usepackage{amsthm}
\usepackage{amsfonts}
\usepackage{graphicx,psfrag,epsf}
\usepackage{enumerate}
\usepackage{natbib}
\usepackage{url} 
\usepackage{subcaption}
\usepackage{algorithm}
\usepackage{algpseudocode}
\usepackage{float} 
\usepackage{pdflscape} 
\usepackage{booktabs}
\usepackage{caption}   
\usepackage{multicol, multirow}
\usepackage{array}
\usepackage{hyperref}
\usepackage{tikz}
\usetikzlibrary{positioning}
\usepackage{makecell}

\newcommand{\blind}{0}

\newcommand{\X}{{\mathbf X}_{i}}
\newcommand{\Z}{Z_{i}}
\newcommand{\Y}{Y_{i}}
\def\Xv{\mathbf X}
\def\Zv{\mathbf Z}
\def\xv{\mathbf x}

\newtheorem{theorem}{Theorem}
\newtheorem{lemma}{Lemma}

\newtheorem{prop}{Proposition}
\newtheorem{definition}{Definition}
\numberwithin{remark}{prop}

\begin{document}

\def\spacingset#1{\renewcommand{\baselinestretch}%
{#1}\small\normalsize} \spacingset{1}


\if0\blind
{
  \title{\bf Mixing Samples to Address Weak Overlap in Causal Inference}
  \author{Jaehyuk Jang, Suehyun Kim, and Kwonsang Lee\\
    Department of Statistics, Seoul National University \date{}} \maketitle
} \fi

\if1\blind
{
  \bigskip
  \bigskip
  \bigskip
  \begin{center}
    {\LARGE\bf Title}
\end{center}
  \medskip
} \fi

\bigskip
\begin{abstract}
In observational studies, the assumption of sufficient overlap (positivity) is fundamental for the identification and estimation of causal effects. Failing to account for this assumption yields inaccurate and potentially infeasible estimators. To address this issue, we introduce a simple yet novel approach, \textit{mixing}, which mitigates overlap violations by constructing a synthetic treated group that combines treated and control units. Our strategy offers three key advantages. First, it improves the accuracy of the estimator by preserving unbiasedness while reducing variance. The benefit is particularly significant in settings with weak overlap, though the method remains effective regardless of the overlap level. This phenomenon results from the shrinkage of propensity scores in the mixed sample, which enhances robustness to poor overlap. Second, it enables direct estimation of the target estimand without discarding extreme observations or modifying the target population, thus facilitating a straightforward interpretation of the results. Third, the mixing approach is highly adaptable to various weighting schemes, including contemporary methods such as entropy balancing. The estimation of the Mixed IPW (MIPW) estimator is done via M-estimation, and the method extends to a broader class of weighting estimators through a resampling algorithm. We illustrate the mixing approach through extensive simulation studies and provide practical guidance with a real-data analysis.
\end{abstract}

\noindent%

\vfill

\newpage
\spacingset{1.45} 

\section{Introduction}
\label{sec1}

Causal inference in observational studies \citep{cochran1965planning} has rapidly caught the eye of diverse applied fields such as epidemiology, political science, sociology and economics since the introduction of the causal framework \citep{rubin1974estimating}. The foundation of the propensity score established by \cite{rosenbaum1983} proliferated various statistical methodologies to estimate the treatment effect by mimicking observational data in randomized experimental data \citep{rubin2007design, imai2008misunderstandings}. The key technique is to control the imbalance of confounders between the treatment variable and the outcome variable through matching \citep{stuart2010matching, rosenbaum2002overt, rosenbaum1989optimal}, stratification \citep{rosenbaum1983, rosenbaum1984reducing} and weighting \citep{hernan2024causal, robins1994estimation}.

However, due to the heavy reliance of observational studies on a pair of key assumptions, \textit{ strong ignorability} \citep{rosenbaum1983}, many causal inference methodologies struggle from its violations. Whereas tools such as sensitivity analysis \citep{rosenbaum2004design} have been devised to account for hidden biases due to unmeasured confounders, causal inference still faces prominent issues regarding the lack of overlap in many situations. This problem is particularly common in the propensity score literature, which may result from the choice of covariates included in the propensity score model (may lead to model misspecification) \citep{matsouaka2024causal} or the inclusion of high-dimensional covariates \citep{d2021overlap}.  In particular, the previous matching literature identified lack of overlap as a key obstacle \citep{visconti2018handling, stuart2010matching}. Moreover, past works on another popular method, weighting, highlighted the severity of the assumption violation by showing the decay of convergence rate \citep{khan2010irregular, hong2020inference}.

Various remedies have been suggested to address the issue of weak overlap, including restricting the targeting population to the population with common support by trimming or truncating \citep{crump2009dealing, lee2011weight}. Another strategy is to estimate a different causal effect that does not heavily rely on, or rely at all on, the overlap assumption by modifying the target estimand\citep{li2018balancing, kennedy2019nonparametric}. However, such methods alter the target population, raising concerns about the validity of such estimators for the original population of interest. Some weighting methods, although not directly focused on overlap violation, were developed to protect inference from extreme weights arising from model misspecification \citep{ben2021balancing, hainmueller2012entropy, imai2014covariate, zubizarreta2015stable}, and have only been validated empirically. Their primary role to prevent bias from model misspecification may not be effective in the presence of weak overlap. Furthermore, the optimization involved in their estimation may become infeasible under such circumstances.

In this article, we propose a simple yet novel approach to address weak overlap by \textit{mixing} the treated and control populations. Specifically, we define a synthetic population --- referred to as the \textit{mixed population} --- as a mixture of the original treated and control populations, and perform causal estimation within this mixed population. This approach alleviates the weak overlap problem in the synthetic population, enabling us to recover more stabilized weights for the original population. Consequently, we introduce the Mixing Inverse Probability Weighting (MIPW) estimator, which substitutes the propensity scores in the standard Inverse Probability Weighting (IPW) estimator with the adjusted mixed weights, and discuss its theoretical properties. Although we mainly focus on mixing in propensity score-based methods, the mixing approach is not confined to the IPW framework. Mixing can be extended to a modern class of weighting methods --- \textit{balancing methods} --- a nonparametric weight estimation method \citep{ben2021balancing} and this is illustrated through a variation of resampling algorithm.

The proposed methodology addresses the limitations of previous approaches for handling weak overlap in the following aspects. First, the mixing estimator enhances the efficiency of current estimators while preserving unbiasedness. This improvement is due to the shrinkage of weights during the mixing process, which mitigates the impact of extreme weights, thereby reducing the variance of the estimator. Another significant benefit is that the mixing estimator targets the true population of the causal estimand, even in the presence of heterogeneous treatment effects. This offers an advantage over methods such as trimming \citep{crump2009dealing} or estimation of the \textit{Average Treatment Effect in the Overlap population (ATO)} \citep{li2018balancing}, which alter the target population to facilitate the estimation procedure. Furthermore, our approach can be easily extended to existing weighting methods by establishing a simple relationship between the original and mixed weights, allowing for an upgrade in existing weighting techniques. While we present an extension of Entropy Balancing \citep{hainmueller2012entropy} in this paper, the mixing approach can be readily applied to other balancing approaches \citep{ben2021balancing}, including Covariate Balancing Propensity Score (CBPS) \citep{imai2014covariate}. 

The remainder of this paper is organized as follows. Section 2 introduces the notation and outlines the problem of estimating causal estimands in the context of empirical weak overlap. In Section 3, we propose the mixing approach by introducing the mixing distribution, along with offering two practical algorithms for causal estimation. In Section 4, a simulation study is conducted to demonstrate the performance of the weighting estimators combined with mixing under various overlap conditions. We then apply the proposed methods to a real-world example in Section 5. Finally, Section 6 concludes the study by summarizing the key takeaways and suggesting a direction for future research. Proofs and additional visualizations are provided in the Supplementary Materials.

\section{Preliminaries}
\label{sec2}

\subsection{Notations and Problem Setup}
\label{sec2.1}

We start with introducing basic notations and causal identification assumptions. In this paper, consider a binary treatment variable; $\Z = 1$ if individual $i$ is treated and $Z_i=0$ otherwise. From the potential outcome framework \citep{rubin1974estimating}, define $Y_i(1)$ as the outcome that would be observed if individual $i$ is treated, and $Y_i(0)$ as the outcome that would be observed if not treated. Consider a dataset $(Y_i(1),Y_i(0),\Z,\X), \; i=1,\dots,N$ where $(Y_i(1), Y_i(0))$ represents the real-valued potential outcome variables supported on $\mathcal{Y}\subseteq \mathbb{R}$ and $\X \in \mathcal{X}\subseteq \mathbb{R}^d$ is a $d$-dimensional covariate vector with support $\mathcal{X}$. Since we cannot observe both of the potential outcomes simultaneously and observe $Y_i=\Z Y_i(1) + (1-\Z) Y_i(0)$ only \citep{holland1986statistics}, the observed dataset $(Y_i, Z_i, \X)$ is considered for analysis.

In order to identify causal effects such as the average treatment effect ($ATE=E[Y(1)-Y(0)]$) or the average treatment effect on the treated ($ATT=E[Y(1)-Y(0)\mid Z=1]$) from observed data, the \textit{strong ignorability} assumptions must be satisfied. Strong ignorability comprises two key conditions: (1) unconfoundedness and (2) strict overlap of the propensity score $e(\xv)$, as defined by Rosenbaum and Rubin (\citeyear{rosenbaum1983}).

\begin{gather}
    (Y(1),Y(0)) \perp \!\!\! \perp Z \mid \Xv\\
    0<e(\xv)=P(Z=1 \mid \Xv = \xv) < 1
\end{gather}

However, these assumptions are not guaranteed to hold in the context of observational studies. While a substantial body of literature focuses on estimating the ATE or ATT, the validation of identification assumptions is often overlooked in practice. Numerous studies emphasize that the performance of existing estimators, in terms of both bias and variance, is highly sensitive to the validity of these conditions. In fact, the two assumptions oppose each other --- a high-dimensional covariate space may increase the plausibility of unconfoundedness (Assumption (1)), but simultaneously undermine the level of overlap (Assumption (2)), as noted by \cite{d2021overlap}.

Our discussion focuses on the violation of Assumption (2), which concerns limited overlap. Assumption (2) ensures that an observation can potentially belong to both the treated and control groups, regardless of covariate values. This condition enables the imputation of unobserved potential outcomes using observations with similar propensity scores or weighted pseudo-outcomes. Thus, sufficient overlap in the propensity score distributions of the treated and control groups is essential; large discrepancies between these distributions lead to unstable comparisons. In extreme cases, where $e(\xv) = 0$ or $e(\xv) = 1$, Assumption (2) is entirely violated.

Violations of overlap can be classified into two types: \textit{structural} and \textit{random} (\cite{hernan2024causal}, Section 12). Structural violation occurs when there exist certain individuals that cannot be either treated or controlled almost surely, implying non-overlapping strata in the propensity score support. In such cases, the violation is deterministic and thus the identification of causal effects is infeasible. On the other hand, random violation arises from sampling variability, resulting in empirical non-overlap. In this paper, we restrict our problem to random violations, assuming that the identification of the ATE and ATT remains feasible.

\subsection{Enhancing Causal Estimation Through Sufficient Overlap}
\label{sec2.2}

Recognizing that the degree of overlap is critical in estimation, we ask a related question: `Can we improve causal estimation if the empirical overlap of the propensity scores increases?' A popular framework proposed by \cite{li2018balancing}, \textit{balancing weights} offers a foundation for exploring this curiosity. Consider the standard propensity score weighting estimator of the ATT.

\begin{align}
    \hat{\tau}_{IPW} = \frac{\sum_i \Z Y_i}{\sum_i \Z} - \frac{\sum_i\frac{e}{1-e}(\X)(1-\Z) Y_i}{\sum_i\frac{e}{1-e}(\X)(1-\Z)}.
\end{align}
where propensity score is modeled as $e(\X)=e(\X;\beta)$ such as the logistic model.

Within the framework, the weighting estimator $\hat{\tau}_{IPW}$ is a special case of the \textit{sample estimators of the weighted average treatment effect (WATE)} targeting the treatment effect of a subpopulation; the treated group. The proof of the theorems in the paper hint several finite-sample properties that we can relate with the degree of empirical overlap. Theorem 1 in \cite{li2018balancing} states that $\hat{\tau}_{IPW}$ is a consistent estimator of $ATT$, which requires the propensity score model to be well-fitted; otherwise the estimation would be biased. Moreover, the proof of Theorem 2 formulates a general form of the conditional variance of $\hat{\tau}_{IPW}$ given the samples $(\Xv, \Zv) = \{(\X, \Z)\}_{i=1}^n$ when the outcome variable is the only random variable.

\begin{align}
    V\left[\hat{\tau_{IPW}}\mid \Xv, \Zv \right] = \frac{\sum_i \Z v_1(\X)}{\left(\sum_i \Z\right)^2} + \frac{\sum_i \left(\frac{e(\X)}{1-e(\X)}\right)^2(1-\Z)v_0(\X)}{\left(\sum_i \frac{e(\X)}{1-e(\X)}(1-\Z)\right)^2}
\end{align}
where $v_z(\X)=V[\Y(z)\mid \X]$. It follows directly from the equation that even if the propensity scores are correctly modeled, the finite-sample performance would deteriorate severely when there is control unit with unusually high propensity score i.e. $e(\X)\approx 1, i:\Z=0$.

Now imagine two sets of random samples $A:\{(\Y,\Z,\X)\}_{i=1}^n, B:\{(\Y^*, \Z^*, \X^*)\}_{i=1}^n$ from the same joint distribution of $(Y,Z,\Xv)$ but differ in the degree of overlap of propensity scores due to sampling variability. Suppose the former is poorly overlapped relative to the latter.

Compared to dataset $B$, the causal estimation with samples $A$ is vulnerable in two different ways. First, separable issue is more likely, especially when the sample size is small \citep{schaefer1983bias}, resulting in a poorly fitted propensity score model that may invalidate the condition of the consistency of the weighting estimator. Solutions, including penalized methods exist for logistic models \citep{heinze2002solution, firth1993bias}, but they can still induce biased result from common modeling issue. Therefore, the ATT estimation is unreliable due to lack of consistency derived from the first theorem in \cite{li2018balancing}. 

Second, even if the propensity scores are well estimated, extreme weights are still hard to avoid. When there are enough data points to bypass complete separability, it means that there are a portion of control units with large propensity scores compared to the rest of the group, thereby creating extreme values of $e(\X)/(1-e(\X))$. As previously noted, these extreme weights are a primary source of large finite-sample variance, consistent with the findings of \cite{busso2014new}.

On the other hand, since dataset $B$ exhibits better overlap, the causal estimation there is relatively more stable. This motivates us to build a strategy to manipulate a dataset to enhance the overlap of propensity scores, and therefore improve the quality of causal inference, particularly when overlap is weak. In the next section, we propose a simple method to do so by mixing the treated and control units.

\section{Sample Mixing Methods}
\label{sec3}

\subsection{Mixing Strategy}
\label{sec3.1}
Assume that the density of $(Y, Z, \Xv)$ exist, $g$, with respect to the product measure $\mu$ (a product measure of Lebesgue measure for continuous variables and counting measure for categorical variables). Denote the joint density of $(Y,\Xv)$ and the conditional joint density given $Z=z$ as $h$ and $h_z, z=0,1$, respectively. Define $H$ and $H_z$ as the corresponding distributions. Let a non-empty convex set of densities on $\mathcal{Y}\times\mathcal{X}$,

\begin{align*}
    \mathcal{M} = \left\{\theta h_1 + (1-\theta) h_0: 0\leq \theta = \theta(\xv) \leq 1, \forall\xv\in \mathcal{X} \right\}
\end{align*}
where $\theta$ is measurable with respect to $\mu$.

\begin{definition}[Mixed Distribution]\label{def 1}
    Let $Z^*$ be a binary variable with marginal probability, $\pi^*$. Define the joint distribution of $(Y^*,\Xv^*)$ as distribution with density,
    $$
    h^* = h^*_{\pi^*,\theta_1,\theta_0} = \pi^*h^*_1 + (1-\pi^*)h^*_0
    $$
    where
    $$
    h^*_z = \theta_z h_1 + (1-\theta_z)h_0 \in \mathcal{M}, z=0,1
    $$
    are the densities of conditional distribution given $Z^*=z, z=0,1$. We will call its distribution, $H^*$ as ``the mixed distribution of $H$".
\end{definition}

We claim that the mixed distribution, $H^*$ is a synthetic version of $H$ but designed to be easier to balance the covariates. This is gained from the increase in overlap between the control and treated group of the mixed distribution which implies that the propensity score densities of each group to be less skewed toward bound $[0,1]$. As a result, the causal estimation within the mixed distribution leads to a more accurate inference. A potential apprehension is that causal inference in the modified distribution may introduce bias in the treatment effect of interest unlike the theoretical example made in Section \ref{sec2.2}. Fortunately, this concern is unnecessary by Theorem \ref{thm 2}.

\begin{lemma}[Synthetic Propensity Score]\label{lem 1}
Let $e^*(\xv)=P(Z^*=1\mid \Xv^*=\xv), \forall x\in \mathcal{X}$ be the propensity score of the mixed distribution. Then,
    $$\frac{e^*}{1-e^*}(\xv) = \frac{\pi^*}{1-\pi^*}\cdot\frac{\theta_1(\xv)\frac{e}{1-e}(\xv) + (1-\theta_1(\xv))\frac{\pi}{1-\pi}}{\theta_0(\xv)\frac{e}{1-e}(\xv) + (1-\theta_0(\xv))\frac{\pi}{1-\pi}}, \hspace{0.5cm} \forall \xv \in \mathcal{X}$$

We will call the propensity score within the mixed distribution, $e^*$ as the ``synthetic propensity score".
\end{lemma}

\begin{theorem}[Positivity Guarantee of the Mixed Distribution]\label{thm 1}

Consider the the mixed distribution in Def. \ref{def 1}. If $Z^* \stackrel{d}{=} Z$, then $0<e^*(\xv)=P(Z^*=1\mid \Xv^*=\xv)<1,\forall \xv\in\mathcal{X}$.
    
\end{theorem}

\begin{theorem}[Identifiability]\label{thm 2}
    Consider the mixed distribution in Def. \ref{def 1}. If $\theta_1\neq\theta_0$, then all the counterfactual expectations; $$E[Y(z)] \hspace{0.5cm}\text{and}\hspace{0.5cm} E[Y(z)\mid Z = 1-z]$$ for $z=0,1$ are identifiable with the mixed distribution.
\end{theorem}

For an example identification of Theorem \ref{thm 2}, the expectation of the counterfactual potential outcome of the treated, $E\left[Y(0)\mid Z = 1 \right]$, is

\begin{align*}\label{identification 1}
    \frac{1-\pi}{\pi}\left\{E\left[w_0(\Xv^*)Z^*Y^*\right]/\pi^* - E\left[w_1(\Xv^*)(1-Z^*)Y^*\right]/(1-\pi^*)\right\}
\end{align*}
where 

\begin{align*}
    w_z(x) = \frac{\theta_z}{\theta_0-\theta_1}(x)\times\frac{\pi}{1-\pi}\frac{(1-\theta_1(x))\frac{\pi^*}{1-\pi^*}-(1-\theta_0(x))\frac{e^*}{1-e^*}(x)}{\theta_0\frac{e^*}{1-e^*}(x) - \theta_1\frac{\pi^*}{1-\pi^*}}
\end{align*}

With appropriate choices of $\pi^*, \theta_0, \theta_1$, we find that no additional identification assumptions or bias corrections are needed to estimate the treatment effects of interest using the mixed samples by Theorem \ref{thm 2}. In particular, when $\pi^* = \pi$, the propensity score of the mixed distribution is almost surely more robust than the original propensity score by Lemma \ref{lem 1}. This ensures Theorem \ref{thm 1}, which states that the overlap in the mixed distribution is greater than that in the original distribution, thereby leading to more stabilized weights. Thus, the samples from the such mixed distribution are more favorable than the original for treatment effect estimation. Throughout the rest of the paper, we specify a simple but powerful class of mixed distribution as an illustration.

\subsection{ATT Estimation within the Simple Mixed Distribution}
\label{sec3.2}

\begin{definition}[Simple Mixing Strategy]
    We will say $H^*$ is ``simple mixed distribution" when it is defined with $\theta_1$ and $\theta_0$ are set as constants and $Z^*$ as $\pi^* = \pi$ from Definition \ref{def 1}.
\end{definition}

In particular, we will set $\theta_0=0$ and $\theta_1=1-\delta$ for some $0<\delta<1$. Then,

\begin{align}
    h^*_1 &= (1-\delta) h_1 + \delta h_0 \\
    h^*_0 &= h_0
\end{align}

Then, by Lemma \ref{lem 1}, observe that the simple mixed distribution for a fixed $\delta$ is supplemented with the propensity score weights,

\begin{align} \label{mixed ps}
        \frac{e^*}{1-e^*}(\xv) = (1-\delta)\frac{e}{1-e}(\xv) + \delta \frac{\pi}{1-\pi}, \hspace{0.5cm} \forall \xv\in \mathcal{X}
\end{align}

Complementing the theoretical observations made in the general overview of mixing strategy in the previous subsection, formula (\ref{mixed ps}) explicitly demonstrates the increase in overlap within the mixed distribution. As the mixing proportion $\delta$ is scaled from $0$ to $1$, synthetic propensity scores $e^*(\xv)$ degenerate from the original propensity scores $e(\xv)$ to the average of the propensity scores (either original or synthetic), $\pi$. This follows from the fact that $x/(1-x)$ is a one-to-one function on the open interval $(0,1)$. As the propensity scores become increasingly coarsened toward the average, their variation is reduced, leading to increased common support between the control and treated groups. However, it should also be noted that a large $\delta$ may induce a loss of information in exchange of robustness, and therefore, an appropriate choice of $\delta$ is necessary. We advise the reader to exclude the situation where $\delta$ is too close to $1$ throughout the main paper and return to Supplementary Material E where we revisit the issue.

To take advantage of our discovery of the robustness of the synthetic propensity score weights, we derive an $ATT$ estimator, parallel to the standard IPW estimator, using the augmented samples of the mixed and the original; $\{(\Y,\Z,\X,\Y^*,\Z^*,\X^*)\}_{i=1}^N$. We call the following weighting estimator, the ``Mixed IPW (MIPW) estimator".

\begin{align}\label{mipw}
            \hat{\tau}_{MIPW} =  \frac{\sum_i Z_iY_i}{\sum_iZ_i} - \frac{\sum_i\left(\frac{e^*}{1-e^*}(\X^*)-\delta\frac{\pi}{1-\pi}\right)(1-Z^*_i)Y^*_i}{\sum_i\left(\frac{e^*}{1-e^*}(\X^*)-\delta\frac{\pi}{1-\pi}\right)(1-Z^*_i)}
\end{align}

\begin{theorem}[Consistency]\label{thm 3}
        Under the strong ignorability assumptions, $\hat{\tau}_{MIPW}$ is a consistent estimator of $ATT$.
\end{theorem}

There are several noteworthy observations. First, because of the robustness of the synthetic propensity score weight, the MIPW estimator should also be less vulnerable to sampling variation compared to the standard IPW. Next, the model of the synthetic propensity score is modeled based on the initial model assumption of the original propensity score, $e(\Xv)=e(\Xv;\beta)$. Thus, we discover that the consistency of this estimator relies on the same assumptions as the consistency of the standard IPW.

Although the MIPW estimator holds such desirable properties, one critical issue remains. In the real world, we can only observe $\left\{(Y_i,X_i,Z_i)\right\}_{i=1}^N$ sampled from $H$, rather than the augmented dataset. Therefore, a practical implementation is required to compute the MIPW estimator using only the observed data. Two different approaches have been proposed in the literature, each with its own advantages.

One approach is based on M-estimation theory. The large-sample theory underlying this method allows us to compute the MIPW estimator using only the observed dataset, while ensuring asymptotic equivalence to estimation based on the augmented samples. This approach facilitates both finite-sample and large-sample inference for the ATT. However, it lacks the flexibility to accommodate estimation beyond parametric models of the propensity score.

An alternative approach involves generating mixed samples through a modified resampling algorithm that we refer to ``Mixing Algorithm". This method enables indirect observation of mixed samples and allows for ATT estimation using various weighting schemes. In this paper, we implement the mixing strategy within the framework of balancing methods \citep{ben2021balancing}, a broad class of weighting estimators that includes the IPW estimator. However, this added flexibility comes at a cost: large-sample inference is not readily available, and the computational burden for statistical inference is greater than that of the first approach.

In the following two subsections, we provide the details of the implementation methods in order.

\begin{table}[H]
        \centering
        \begin{tabular}{c|c|c}
             & M-estimation & Mixing Algorithm \\
            \hline
            Finite-sample Inference & O & O \\
            Parametric Approach & O & O \\
            Large-sample Inference & O & X \\
            Nonparametric Approach & X & O \\
            Computation Speed & Fast & Slow \\

        \end{tabular}
        \caption{Comparison of Mixing Implementations for Causal Estimation}
        \label{tab:my_label}
\end{table}

\subsection{The M-Estimation-Based Method}
\label{sec3.3}
In this subsection, we show how we can compute the MIPW estimator with the observed dataset alone. We first derive the asymptotic property of the MIPW estimator through M-estimation theory which was solidly reviewed by \cite{stefanski2002calculus} analogous to IPW \citep{lunceford2004stratification}. Consider the joint distribution $J$ of $(Y,Z,\Xv,Y^*,Z^*,\Xv^*)$, where $\Xv,\Xv^*$ are centered for simplicity, and its parameter space $\Theta\subseteq\mathbb{R}^{d+3}$. Suppose we have an i.i.d sample, $\{(\Y,\Z,\X,\Y^*,\Z^*,\Xv^*)\}_{i=1}^N$. Denote $ATT$ as $\tau$ and $\mu(z) = E\left[Y(z)\mid Z=1\right], z=0,1$ for convenience. Assume that the true propensity score is a parametric model, $e(\Xv)=e(\Xv;\beta)$, that is twice differentiable with respect to its parameter and the log likelihood is strictly concave e.g. a logistic regression model. Define the true parameter value $\theta_0 = (\mathbf{\beta}_0, \pi_0, \mu_0(1), \mu_0(0))$ as the unique solution of $E_J[\psi^*(\theta_0)]=0$ where $\psi^*$ is

\begin{align}\label{MIPW_GEE}
    \psi^*(\theta ; Y,Z,\Xv,Y^*,Z^*,\Xv^*)=\begin{pmatrix}
        \frac{Z^*-e^*(\Xv^*;\beta)}{e^*(\Xv^*;\beta)(1-e^*(\Xv^*;\beta))}\nabla_\beta e^*(\Xv^*;\beta)\\
        Z - \pi\\
        Z\left(Y - \mu(1)\right)\\
        \left(\frac{e^*}{1-e^*}(\Xv^*;\beta)-\delta \frac{\pi}{1-\pi}\right)(1-Z^*)\left(Y^* -\mu(0)\right)
    \end{pmatrix}
\end{align}

Then, there exists an M-estimator $\hat{\theta}$ i.e. $\sum_i\psi(\hat{\theta};\Y,\Z,\X,\Y^*,\Z^*,\Xv^*)=0$ and $\hat{\theta}$ satisfies $\hat{\theta}\stackrel{p}{\rightarrow}\theta_0$. Furthermore, since $\psi^*$ is generally smooth under strong ignorability assumptions, $\hat{\theta}$ is asymptotically normal. As a result, we have the asymptotic property of the MIPW estimator, $\hat{\tau}_{MIPW} = \mathbf{c}^T\hat{\theta}$, $\mathbf{c}=(0,0,1,-1)^T$.

\begin{align}\label{MIPW_asymptotic}
    \sqrt{N}^{-1}(\hat{\tau}_{MIPW} - \tau) \stackrel{d}{\rightarrow} N(0, V_{MIPW})
\end{align}
where $V_{MIPW}$ is the asymptotic variance of the MIPW estimator derived through the sandwich variance formula,

\begin{align*}
    V_{MIPW} = \mathbf{c}^T\left(E\left[-\dot{\psi}^*(\theta_0)\right]\right)^{-1}E\left[\psi^*(\theta_0)\psi^*(\theta_0)^T\right]\left(E\left[-\dot{\psi}^*(\theta_0)\right]\right)^{-T}\mathbf{c}
\end{align*}




The M-estimation theory strengthens Theorem \ref{thm 3} by endowing a stronger modeling assumption to the original propensity score. Furthermore, the existence of an equivalent estimating equation—a key concept in M-estimation theory—allows us to address the critical issue of inaccessibility to the mixed samples. According to \cite{stefanski2002calculus}, the estimating equation for a particular estimator is not unique -- many lead to the same quantity of interest. Although different estimating equations may induce different ``bread" and ``meat" of the sandwich variance, it all leads to the same asymptotic variance. We derive an estimating equation that only requires the observed samples, which yields the same property \eqref{MIPW_asymptotic} of the MIPW estimator.

\begin{theorem}[Asymptotic Normality Based on Observed Samples]\label{thm 4}
    Let $\left\{\left(\Y, \Z, \X\right)\right\}_{i=1}^N$ be i.i.d samples from $G$. Define the true parameter value $\theta_0 = (\beta_0,\pi_0,\mu_0(1),\mu_0(0))\in\Theta\subset\mathbb{R}^{d+3}$ as $E_G\left(\psi^{**}(\theta_0)\right)=0$ where $\psi^{**}$ is
    \begin{align*}
        \psi^{**}(\theta; Y, \Xv, Z)=\begin{pmatrix}
        \left\{\frac{1-\delta}{e^*(\Xv;\beta)}Z + \left(\frac{\delta\pi}{(1-\pi)e^*(\Xv;\beta)}-\frac{1}{1-e^*(\Xv;\beta)}\right)(1-Z)\right\}\nabla_\beta e^*(\Xv;\beta)\\
        Z - \pi\\
        ZY - Z\mu(1)\\
        \frac{e(\Xv;\beta)}{1-e(\Xv;\beta)}(1-Z)Y - \frac{e(\Xv;\beta)}{1-e(\Xv;\beta)}(1-Z)\mu(0)
    \end{pmatrix}
    \end{align*}
    for $0<\delta<1$.\footnote{Note that when $\delta=0$, $\psi^{**}$ is equivalent to the estimating equation of IPW. Moreover, the regularity condition of $\psi^{**}$ does not hold when $\delta=1$.} Let $\hat{\theta}$ be the unique solution of $\sum_i\psi^{**}(\hat{\theta};\Y,\Z,\X)=0$ and set $\hat{\tau}_{MIPW} = \mathbf{c}^T\hat{\theta}$. Then, $\hat{\tau}_{MIPW}$ satisfies the property \eqref{MIPW_asymptotic}. 
\end{theorem}

With the new estimating equation, $\psi^{**}$, we can make statistical inference, in terms of both finite- and large-sample based, on the ATT with the MIPW estimator only with the observed dataset while remaining asymptotically equivalent to the estimation based on the augmented dataset. According to our numerical studies, the MIPW estimator outperforms IPW with respect to efficiency with or without weak overlap, unless $\delta$ is too large.

\subsection{The Mixing Algorithm-Based Method}
\label{sec3.4}

In the previous section, we computed the MIPW estimator through a method that heavily relies on M-estimation theory but not all causal inference methodologies depend on it. To extend our mixing idea to broader weighting methodologies, in case where the propensity scores can be modeled without strict restriction, we face the need to actually sample data of the mixed distribution. Once we have access to mixed samples, we can not only apply mixing to the IPW framework using equation \eqref{MIPW_GEE}, but also employ completely nonparametric approaches such as \textit{balancing methods} \citep{ben2021balancing}. In this subsection, we establish a modified resampling algorithm that we refer to as the \textit{mixing algorithm} and illustrate the traditional propensity score weighting procedure that mimics the result of previous section for a simple example. Then, we expand the usage to a more comprehensive and modern technique.


To mimic random sampling of the mixed distribution (5), (6) for a fixed $\delta\in(0,1)$, we resample from the original treated and control dataset with a ratio of $1-\delta:\delta$ and label it as the mixed treated dataset. On the other hand, we simply resample the original control dataset and label it as the mixed control dataset. We repeat this procedure $M$ times to attain numerous datasets. The specific resampling algorithm is illustrated in the Algorithm \ref{alg 1}.

Once $M$ mixed dataset are generated from the algorithm, we estimate maximum likelihood estimators of the log-likelihood of synthetic propensity scores for each samples i.e. $\hat{\beta}^{(m)}=\underset{\beta\in\Theta}{\arg\max}\sum_i \ell^{*}(\beta^{(m)};\Xv^{*,(m)}_i,Z_i), m=1,\dots,M$. Then, to control the variability of resampling, like bagging \citep{breiman1996bagging}, we derive the weight by appropriately adjusting and averaging, $W^M_i=\frac{1}{M}\sum_{m=1}^M w^{*,(m)}_i$ where $w^{*,(m)}_i= \left(\frac{e^*}{1-e^*}(\Xv^{*,(m)}_i;\hat{\beta}^{(m)})-\delta \frac{\hat{\pi}}{1-\hat{\pi}}\right), i=1,\dots,N$ for each $m=1,\dots,M$. Finally, we calculate the estimate of the ATT and denote it as $\hat{\tau}_{MIPW.M}$ as \eqref{mipw.m}. To distinguish from the $MIPW$ estimator derived from the M-estimation-based approach, we will call this as the ``$MIPW.M$ estimator" in this paper.

\begin{align}\label{mipw.m}
    \hat{\tau}_{MIPW.M} = \frac{\sum_i \Z Y_i}{\sum_i \Z} - \frac{\sum_iW^M_i(1-\Z) Y_i}{\sum_iW^M_i(1-\Z)}
\end{align}

\begin{algorithm}[H]
\caption{Mixing Algorithm}
\label{alg 1}
\begin{algorithmic}
    \State \textbf{Inputs:} $\{(Y_i,\Xv_i,Z_i)\}_{i=1}^N$, $\delta$, $M$
    \State \textbf{Output:} $\{(Y^{*,(m)}_i,\Xv^{*,(m)}_i),Z_i\}_{i=1}^{N}, m = 1,\dots,M$ \Comment{Mixed dataset}
    \State $N_t \gets \sum_i Z_i$
    \For{$m = 1$ to $M$}
        \State $\{(Y^{*,(m)}_i,\Xv^{*,(m)}_i,Z_i)\}_{i=1}^N\gets$ copy of $\{(Y_i,\Xv_i,Z_i)\}_{i=1}^N$
        \State $I^*_j \stackrel{i.i.d}{\sim} Ber(\delta)$ for $i\in\{Z_i=1\}$
        \State $N^* \gets \sum_i I^*_i$
        \State $N_c^* \gets N_t - N^*$
        \State $(Y^{*,(m)}_i,\Xv^{*,(m)}_i, Z_i=1) \gets$ sample $N^*$ without replacement from $\{(Y_i,\Xv_i)\}_{\{i:I^*_i=1\}}$
        \State $(Y^{*,(m)}_i,\Xv^{*,(m)}_i, Z_i=1) \gets$ sample $N_c^*$ with replacement from $\{(Y_i,\Xv_i)\}_{\{i:Z_i=0\}}$
    \EndFor
    \State \Return $\{(Y^{*,(m)}_i,\Xv^{*,(m)}_i),Z_i\}_{i=1}^{N}$ for $m = 1,\dots,M$
\end{algorithmic}
\end{algorithm}

The reason we do not average the whole weighted mean i.e. $\frac{1}{M}\sum_m\frac{\sum_i w^{*,(m)}_i(1-\Z^{*,(m)})\Y^{*,(m)}}{\sum_i w^{*,(m)}_i(1-\Z^{*,(m)})}$ is to reduce the joint sampling variability of $(Y,Z, w^*)$. Simply plugging-in the averaged weight, $W^M_i$ is reasonable since the algorithm is designed to $w^{*,(m)}_i(1-\Z^{*,(m)})Y^{*,(m)}=w^{*,(m)}_i(1-\Z)\Y,\forall i,m$. We recommend the reader to set $M$ above $200$ for accurate weight estimation. The plot in the Supplementary Material B, we demonstrate how well $MIPW.M$ estimator complements to $MIPW$ estimator in the same simulation setting as in subsection \ref{sec4.1}.

One big advantage of mixing algorithm is that it enables us to extend our approach to a variety of existing methodologies such as \textit{balancing methods}\citep{ben2021balancing, hainmueller2012entropy, imai2014covariate, zubizarreta2015stable}. \cite{ben2021balancing} gives an excellent introduction about balancing approach by comparing it with the conventional modeling approach. The following proposition guides us on how to adjust the derived balancing weight of the mixed samples that can balance the original samples.

\begin{prop}[Contribution of Balancing Mixed Covariates]\label{prop 3}
    For all bounded $f$, suppose we have $W^*$ such that
    \begin{align*}
        E\left[Zf(\Xv^*)\right] = E\left[W^*(\Xv^*)(1-Z)f(\Xv^*)\right]
    \end{align*}
    Let $W=\left(W^*-\delta w_\pi\right)/\left(1-\delta\right)$. Then,
    \begin{align*}
        E\left[Zf(\Xv)\right] = E\left[W(\Xv)(1-Z)f(\Xv)\right]
    \end{align*}
\end{prop}

Observe that $W^*$ is a one-to-one representation of synthetic propensity score weights, $e^*/(1-e^*)$ and likewise for $W$. Derived from equation \eqref{mixed ps}, $W^*$ should be a more safe statistic to estimate under limited overlap. 

In practice, we estimate the balancing weight of the observed samples, $W$, by looking for $w^{*,(m)}_i=w^{*}(\Xv^{*,(m)}_i),\forall i$ for each $m=1,\dots,M$ mixed mixed dataset, such that 

\begin{align}
    \frac{1}{n}\sum_i Z^{*,(m)}_i f(\Xv^{*,(m)}_i) \approx \frac{1}{n}\sum_i w^{*,(m)}_i(1-Z^{*,(m)}_i)f(\Xv^{*,(m)}_i)
\end{align}
and then calculate $W_i = \frac{1}{M}\sum_m w^{(m)}_i$ where $w^{(m)}_i=\left(w^{*,(m)}_i-\delta\frac{\hat{\pi}}{1-\hat{\pi}}\right)/(1-\delta),i =1,\dots,N$. Then plug-in in the equation \eqref{mipw.m}. We report the results from the applying mixing algorithm to the balancing approaches including \textit{Entropy Balancing (EB)} \citep{hainmueller2012entropy} in subsection \ref{sec4.2} and the \textit{just-identified Covariate Balancing Propensity Scores (CBPS)} \citep{imai2014covariate} in the Supplementary Material C. 

\section{Simulation Studies}
\label{sec4}

\subsection{Application onto IPW via M-Estimation-Based Method}
\label{sec4.1}

In this subsection, we conduct a simulation study to visualize how mixing improves IPW in terms of efficiency in $ATT$ estimation within different overlap conditions. We will examine the enhancement of the mixing with respect to the change of $\delta$. Furthermore, we set a benchmark for efficiency, the \textit{Overlap Weighting (OW)} estimator \citep{li2018balancing} of the $ATO$ to numerically assess whether the efficiency gains from mixing can approach the minimum estimation variability.

\begin{align*}
    ATO = \frac{E\left[(1-e(X))ZY\right] - E\left[e(X)(1-Z)Y\right]}{E\left[e(x)(1-e(x))\right]}
\end{align*}

Proposed by \cite{li2018balancing}, $ATO$ is a causal estimand specifically developed to address weak overlap estimated by the $OW$ estimator. It represents the treatment effect in a subpopulation for which the average treatment effect can be estimated with the smallest variance in certain condition. We design our study so that no causal estimators can beat the $OW$ estimator in terms of variability. We also construct the simulation to be $ATT=ATO$ for fair comparison.

We generate a 5-dimensional covariates from the multivariate standard normal distribution i.e. $\Xv=(X_1,X_2,X_3,X_4,X_5) \sim N_5(\mathbf{0}, I_5)$ with sample size $N=1000$. Assuming the logistic regression model for the true propensity score i.e. $e(\Xv;\beta) = \{1+\exp(-\Xv^T\beta)\}^{-1}$, we control three different overlap conditions by : (1) strong overlap: $\beta = (-0.5, 0.5, -0.5, 0.5, 0.5, 0.5)$, (2) moderate overlap: $\beta = (-1, 1, -1, 0.5, -0.5, 0.5)$, (3) weak overlap: $\beta = (-2, 2, -2, 1, 0, 0)$. We random sample the treatment variable $Z \sim Bernoulli(e(\Xv;\beta))$.  Finally, the observed outcome variable is generated via $Y\sim N(E[Y\mid X,Z],1)$ where the conditional mean is a linear model i.e. $E[Y\mid X,Z] = \Xv \gamma + \tau Z, \gamma = (2,2,2,0,1,-1)$ and $\tau=1$. We repeat the data generation for each $\delta$ a sequence from $0.05$ to $0.95$ by an interval $0.05$ with $3000$ replications in every overlap setting.

For each simulated data in the $3000$ replications, we estimate the standard IPW estimator, the OW estimator and the MIPW estimator with the M-estimation-based method. For each estimator, we also compute the Huber-White's robust standard error estimates using the variance sandwich formula \cite{stefanski2002calculus}. Then, we record the average of the point estimates to capture the finite-sample bias. Moreover, we calculate the standard deviation estimates -- capturing the efficiency of finite-sample inference and the average of the Huber-White's robust standard error estimates -- representing the efficiency of large-sample inference.

\begin{figure}[h]
    \centering
    \includegraphics[width=\textwidth]{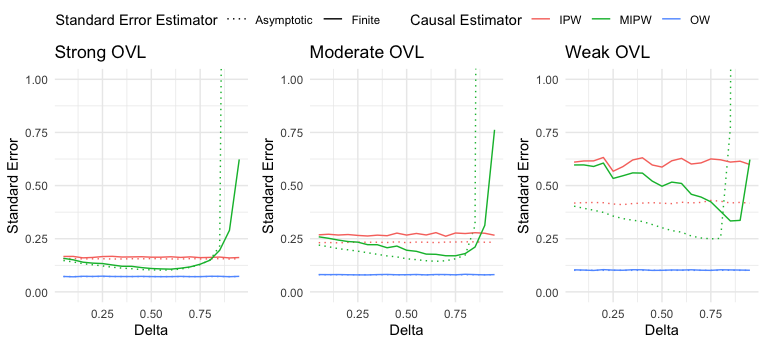}
    \caption{Monte Carlo simulated result of the efficiency of the estimators: (solid lines) the standard deviation estimates and (dotted lines) Huber-White's robust standard error estimates.}
    \label{fig: 1}
\end{figure}

The result of the Monte Carlo $ATT$, $ATO$ estimates and standard deviation estimates of each estimators is in Table \ref{tab: ipw vs mipw vs ow}. Figure \ref{fig: 1} is a plot of the standard deviation estimates and the Huber-White's robust standard error estimates on every $\delta$. 

In Figure \ref{fig: 1}, note that both the red and blue lines, obviously, stay flat since the $IPW$ and the $OW$ estimators are not variant to $\delta$. However, observe that the red solid lines get more wiggly as the overlap weakens (from left to right) indicating the increase in finite-sample variability. Furthermore, both standard error estimates of $IPW$ increase to larger values and the gap between the solid and dotted line widens. Complying with \cite{hong2020inference} in $ATE$ estimation, this numerical observation illustrates the convergence rate of $IPW$ is slower as the overlap violation is more severe. Therefore, estimating $ATT$ via $IPW$ should be waived in limited overlap. In contrast, the two blue lines are remain steady and well-aligned to each other in every condition, indicating that the OW estimator is a perfect golden standard of efficiency in our simulation study.

\begin{landscape}
\begin{table}[ht]
\centering
\resizebox{\linewidth}{!}{%
\begin{tabular}{c|ccc|ccc|ccc}
\multirow{2}{*}{$\delta$} & \multicolumn{3}{c|}{\underline{\textbf{Strong}}} & \multicolumn{3}{c|}{\underline{\textbf{Moderate}}} & \multicolumn{3}{c}{\underline{\textbf{Weak}}} \\
         & IPW & MIPW & OW & IPW & MIPW & OW & IPW & MIPW & OW \\
\midrule
0.05 & 1.001 (0.167) & 1.001 (0.158) & 0.999 (0.072) & 0.996 (0.268) & 0.996 (0.259) & 1.000 (0.081) & 0.995 (0.610) & 0.996 (0.597) & 1.000 (0.104) \\
0.10 & 0.998 (0.167) & 0.998 (0.151) & 1.000 (0.071) & 0.998 (0.271) & 0.998 (0.252) & 1.001 (0.081) & 0.995 (0.616) & 0.996 (0.597) & 1.000 (0.102) \\
0.15 & 0.999 (0.160) & 1.000 (0.140) & 1.003 (0.073) & 0.986 (0.267) & 0.986 (0.244) & 1.002 (0.081) & 1.004 (0.616) & 1.003 (0.590) & 0.999 (0.101) \\
0.20 & 0.999 (0.162) & 1.000 (0.135) & 1.001 (0.073) & 0.989 (0.270) & 0.990 (0.237) & 1.001 (0.081) & 1.000 (0.632) & 1.002 (0.606) & 0.999 (0.105) \\
0.25 & 1.003 (0.166) & 1.003 (0.133) & 1.000 (0.074) & 0.995 (0.266) & 0.993 (0.234) & 0.998 (0.080) & 0.977 (0.567) & 0.977 (0.534) & 1.000 (0.102) \\
0.30 & 0.999 (0.167) & 1.001 (0.127) & 1.000 (0.072) & 0.996 (0.263) & 0.996 (0.222) & 1.000 (0.080) & 0.995 (0.589) & 0.998 (0.546) & 0.999 (0.102) \\
0.35 & 1.006 (0.164) & 1.002 (0.120) & 1.000 (0.072) & 0.998 (0.267) & 0.997 (0.222) & 1.002 (0.081) & 0.992 (0.620) & 0.997 (0.560) & 1.001 (0.104) \\
0.40 & 1.005 (0.164) & 1.002 (0.120) & 1.000 (0.072) & 0.992 (0.264) & 0.991 (0.208) & 1.002 (0.082) & 1.004 (0.631) & 0.998 (0.559) & 1.001 (0.104) \\
0.45 & 0.997 (0.165) & 0.999 (0.114) & 1.000 (0.073) & 0.999 (0.276) & 0.997 (0.215) & 1.000 (0.080) & 0.994 (0.597) & 1.002 (0.521) & 1.000 (0.101) \\
0.50 & 1.000 (0.163) & 0.999 (0.110) & 0.998 (0.072) & 0.996 (0.267) & 0.996 (0.196) & 1.001 (0.081) & 0.992 (0.587) & 1.002 (0.497) & 1.001 (0.101) \\
0.55 & 0.997 (0.163) & 1.000 (0.108) & 1.000 (0.071) & 0.994 (0.275) & 0.994 (0.191) & 0.999 (0.082) & 1.007 (0.617) & 1.004 (0.516) & 1.002 (0.103) \\
0.60 & 1.002 (0.165) & 0.999 (0.107) & 0.999 (0.072) & 0.995 (0.268) & 0.993 (0.178) & 1.001 (0.080) & 0.992 (0.628) & 0.999 (0.510) & 0.997 (0.102) \\
0.65 & 0.997 (0.162) & 1.001 (0.111) & 0.999 (0.073) & 0.992 (0.279) & 0.996 (0.177) & 1.001 (0.082) & 1.016 (0.601) & 1.007 (0.459) & 1.000 (0.104) \\
0.70 & 1.003 (0.165) & 1.003 (0.118) & 1.001 (0.072) & 0.996 (0.262) & 0.999 (0.170) & 1.002 (0.081) & 1.003 (0.607) & 0.998 (0.446) & 1.001 (0.102) \\
0.75 & 1.004 (0.161) & 1.000 (0.130) & 1.000 (0.072) & 0.994 (0.277) & 0.999 (0.170) & 0.998 (0.080) & 1.004 (0.625) & 1.003 (0.424) & 1.002 (0.101) \\
0.80 & 0.997 (0.162) & 1.001 (0.149) & 1.000 (0.073) & 0.996 (0.274) & 1.010 (0.180) & 1.002 (0.083) & 1.004 (0.621) & 1.007 (0.379) & 1.001 (0.104) \\
0.85 & 0.997 (0.164) & 1.005 (0.196) & 1.000 (0.073) & 0.986 (0.278) & 1.019 (0.211) & 0.999 (0.081) & 1.009 (0.611) & 0.991 (0.334) & 0.999 (0.103) \\
0.90 & 1.010 (0.160) & 0.998 (0.289) & 1.003 (0.071) & 0.988 (0.276) & 1.052 (0.313) & 0.998 (0.080) & 1.001 (0.615) & 0.995 (0.336) & 1.001 (0.103) \\
0.95 & 1.003 (0.162) & 0.983 (0.623) & 1.000 (0.073) & 0.996 (0.266) & 1.181 (0.762) & 0.997 (0.081) & 0.998 (0.599) & 0.993 (0.622) & 0.996 (0.102) \\
\bottomrule
\end{tabular}
}
\caption{Monte Carlo simulated result of the point estimates and the standard deviation estimates (filled in the parenthesis in each cell) of the IPW, MIPW, OW estimators on each overlap condition.}
\label{tab: ipw vs mipw vs ow}
\end{table}
\end{landscape}

Now consider the green curves, the MIPW estimator. Examine that both finite- and large-sample standard error estimates of the MIPW estimator are convex in $\delta$. We find that unless $\delta$ is too large, in both sample-size perspectives, MIPW estimator performs conspicuously better than the IPW estimator in every settings. What is notable is that we do not see any finite-sample bias trade-off in such cases according to the Table \ref{tab: ipw vs mipw vs ow}. Although, the empirical enhancement of applying mixing to the IPW is dramatic, we see that there is a room for further improvement when applied to stronger methods based on the gap between the trough of the green curves and blue lines.

We have also calculated $MIPW.M$ estimator with $M=200$ on the identical setting. The plots of the simulation study are depicted in Supplementary Material B. Moreover, we recommend the reader to read Supplementary Material E for those wondering about choosing the most appropriate $\delta$ in the real world.

\subsection{Application onto EB via Mixing Algorithm-Based Method}
\label{sec4.2}

Recall the main strengths of the balancing methods include (1) addressing extreme weight problem (2) avoiding propensity score model misspecification. In this subsection, we show that the specialties of balancing techniques can be further amplified when integrated with mixing strategy. In fact, we numerically show that the mixing strategy can complement the balancing methods in limited overlap. 

We will implement mixing to one of the most popular balancing methods, EB, proposed by \cite{hainmueller2012entropy}. The author kindly enlisted the potential limitations where EB may run into problems including the limited degree of overlap that contributes to extreme weights\cite{hainmueller2012entropy}. The paper introduces a refining scheme that trim the large weights throughout the iteration algorithm of weight solution to lower the variance. However, as previously noted, discarding units eventually leads to estimation of a causal effect on unrecognized subpopulation. The mixing algorithm, on the other hand, does not lose the external validity, thereby estimating the target of interest.

We apply mixing into EB in two different scenarios. First, we take exactly the same weak overlap setting from the previous study. Secondly, we modify a well-known experiment proposed by \cite{10.1214/07-STS227} for model misspecification in propensity scores by letting the observed covariates to be irrelevant, $\tilde{\Xv} = (\tilde{X}_1,\tilde{X}_2,\tilde{X}_3,\tilde{X}_4,\tilde{X}_5)=\{\exp(X_1/2),X_2/(1+\exp(X_1))+10,(X_1X_3/25+0.6)^3,(X_1+X_5+20)^2,\sqrt{|X_3-X_5+1|}\}$. We generate the treatment variable through the true propensity scores with the same coefficients in order to assign weak overlap analogously. Benchmarking the rest of the study settings, we chose the coefficients for the observed outcome variable as $\gamma = (-13.7, 27.4, 13.7, 13.7, 13.7, 13.7)$ and the treatment effect $\tau = 210$ which is also designed for the OW estimator to be dominant in terms of efficiency. We again repeat the data generation procedure $3000$ times on each $\delta$ of the same sequence as the previous study.

\begin{figure}
    \centering
    \includegraphics[width=\textwidth]{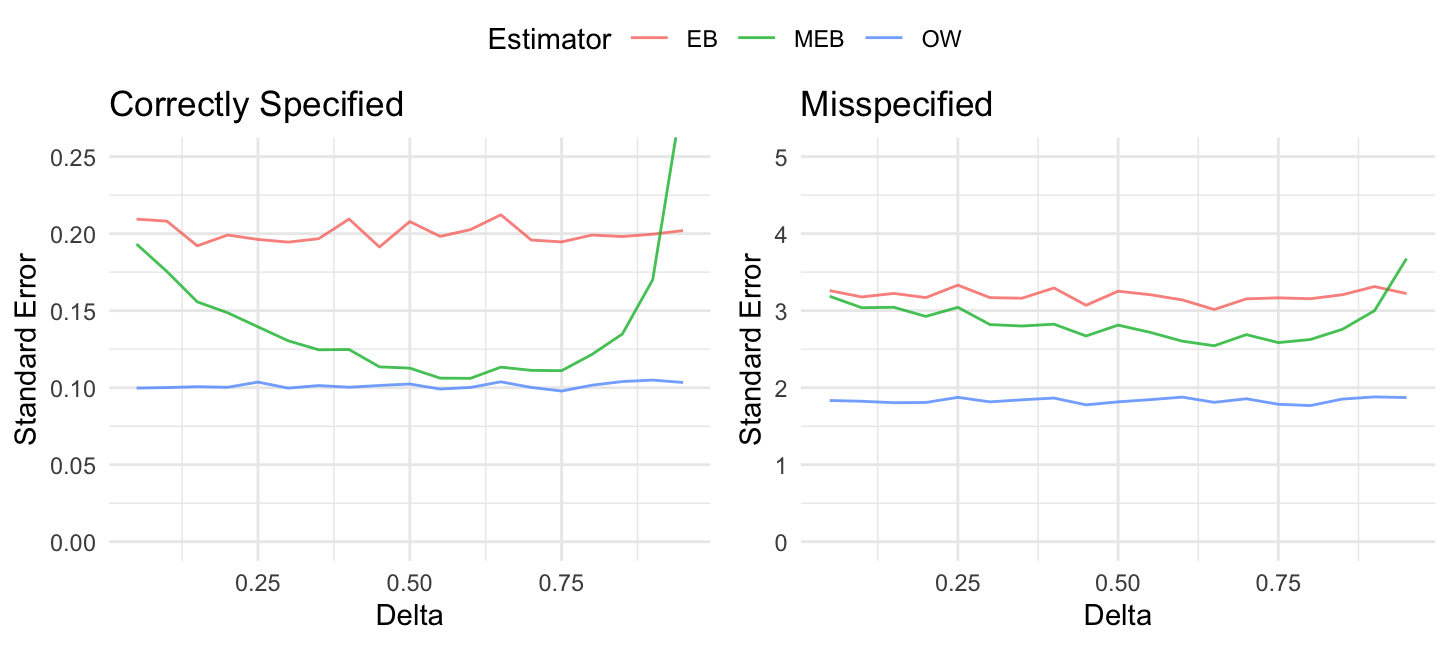}
    \caption{The standard deviation estimates of the estimators EB, Mixing + EB (MEB), OW in weak overlap in case of correct model specification (right) and misspecification (left).}
    \label{fig: 2}
\end{figure}

\begin{figure}
    \centering
    \includegraphics[width=\textwidth]{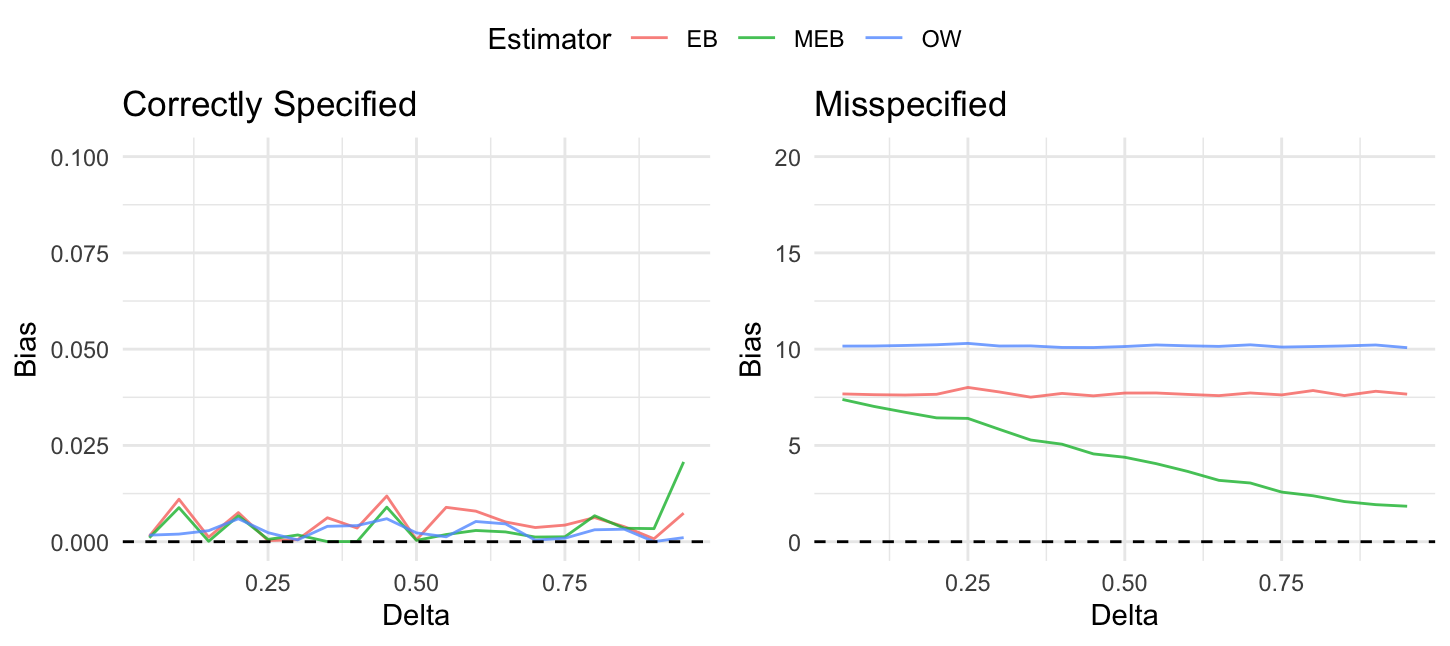}
    \caption{The finite-sample bias of the estimators EB, MEB, OW in weak overlap in case of correct model specification (right) and misspecification (left).}
    \label{fig: 3}
\end{figure}

For each scenario, we calculated two EB estimators one implemented with the original data and the other through the mixing algorithm with $M=200$ along each $\delta$ as in subsection \ref{sec3.4}. We also calculated the OW estimator to check whether the combination of mixing and EB touches the golden standard of efficiency. Since we cannot make large-sample inference through the mixing algorithm-based method, we only record the average point estimates and the standard deviation estimates to observe the finite-sample performance.

The results are depicted in Figure \ref{fig: 2} and \ref{fig: 3}. Regarding the red line in the left panel of Figure \ref{fig: 2}, we see that the entropy balancing improves the weighting method immensely in weak overlap given that the Monte Carlo standard error of the IPW with the conventional propensity score modeling hovered around $0.625$ in a more wiggly manner. Note such gain is further improved when mixing is combined. Moreover, we see that the improvement when $\delta$ is around $0.55 \sim 0.60$ reaches maximum considering the green curve almost touches the blue line which is the benchmark of the efficiency in this simulation study. Similarly, we observe that there is also efficiency gain in case when model is misspecified.

In Figure~\ref{fig: 3}, we observe a more revealing pattern. According to the left panel of Figure~\ref{fig: 3}, there is no bias in any of the estimators when the propensity score model is correctly specified, except for mixing applied to EB with large values of~$\delta$. In contrast, the right panel shows moderate bias in both the EB and OW estimators. In fact, when integrating insights from both panels, we find that neither estimator includes the true treatment effect within its finite-sample $95\%$ confidence interval. The EB estimator partially mitigates the bias from model misspecification, as its bias is no larger than that of OW. However, this is not sufficient.

Interestingly, mixing can compensate for this limitation when paired with an appropriate value of~$\delta$. When $\delta$ is large enough—but not too large—accurate inference with EB becomes possible with the aid of the mixing algorithm. In terms of MSE, we conclude that there is a substantial gain in overall finite-sample performance. We observed a similar result when mixing is applied to another popular balancing method; the \textit{just-identified CBPS} \cite{imai2014covariate} illustrated in Supplementary Material C.

Utilizing the results from the simulation study, we apply mixing onto IPW and EB to reanalyze a famous real world application in the next section.

\section{Real Data Exercise}
\label{sec5}

As one of the ground breaking real-world practice, \cite{hirano2001estimation} reanalyzed the Right heart catheterization (RHC) study\footnote{\cite{connors1996effectiveness} revealed the false delusion of once widely used therapy using propensity score matching method from data collected by \cite{murphy1990support}.} by regression adjustment using the propensity score weights. They selected a large number of covariates to estimate the propensity scores from dataset consisted of $2184$ treated units and $3551$ control units. The researchers selected $56$ covariates out of $72$ to estimate ``better" propensity score for sufficient overlap, hence, succeeding in covariate balancing. Without the tedious covariate selection process, on the other hand, \cite{crump2009dealing} and \cite{li2018balancing} revisited the study using all of the covariates to estimate the propensity score with the logistic regression model. However, instead of estimating the $ATT$ like the previous analyst, they focused on estimating the average causal effect of a subpopulation chosen for efficiency. Regardless, every analyses concluded that applying RHC leads to a higher mortality rate.

Although the overlap is sufficient, we investigate whether mixing can still improve the inference made by the IPW estimator through M-estimation-based method for each $\delta$ from $0.1$ to $0.9$ by an interval $0.1$. To avoid any possible model misspecification, we also estimate the EB and apply the mixing algorithm with $M=200$. To construct the 95\% confidence interval of the $ATT$ estimates, we calculate the bootstrap standard error estimates with $B=2000$. We also calculate the Huber-White's robust standard error estimates of the MIPW. In addition, we estimate the OW for the consistency of our simulation study design.

\begin{figure}[!htb]
   \begin{minipage}{0.5\textwidth}
     \centering
     \includegraphics[width=\linewidth]{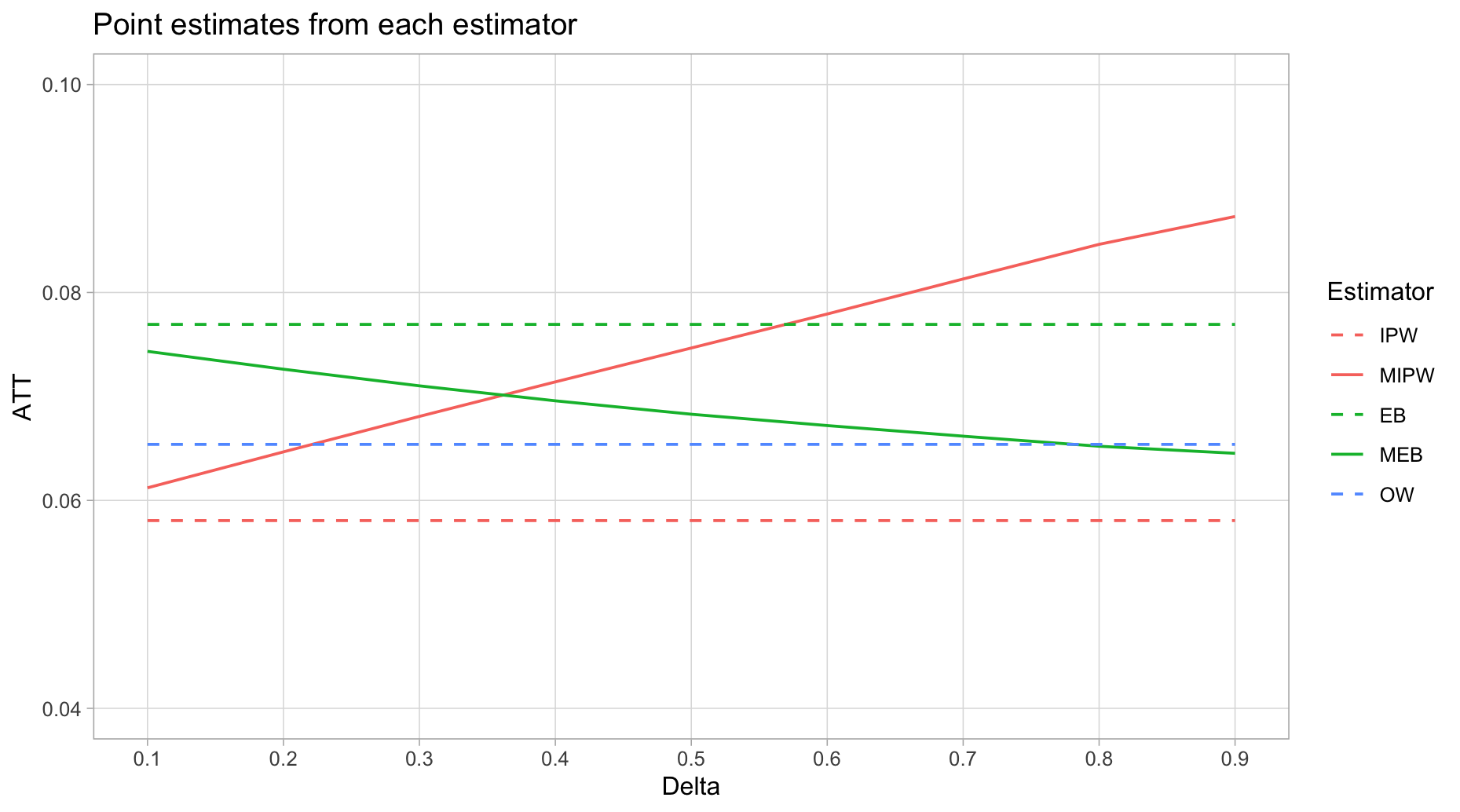}
     \caption{Point Estimates}\label{fig: 5}
   \end{minipage}\hfill
   \begin{minipage}{0.48\textwidth}
     \centering
     \includegraphics[width=\linewidth]{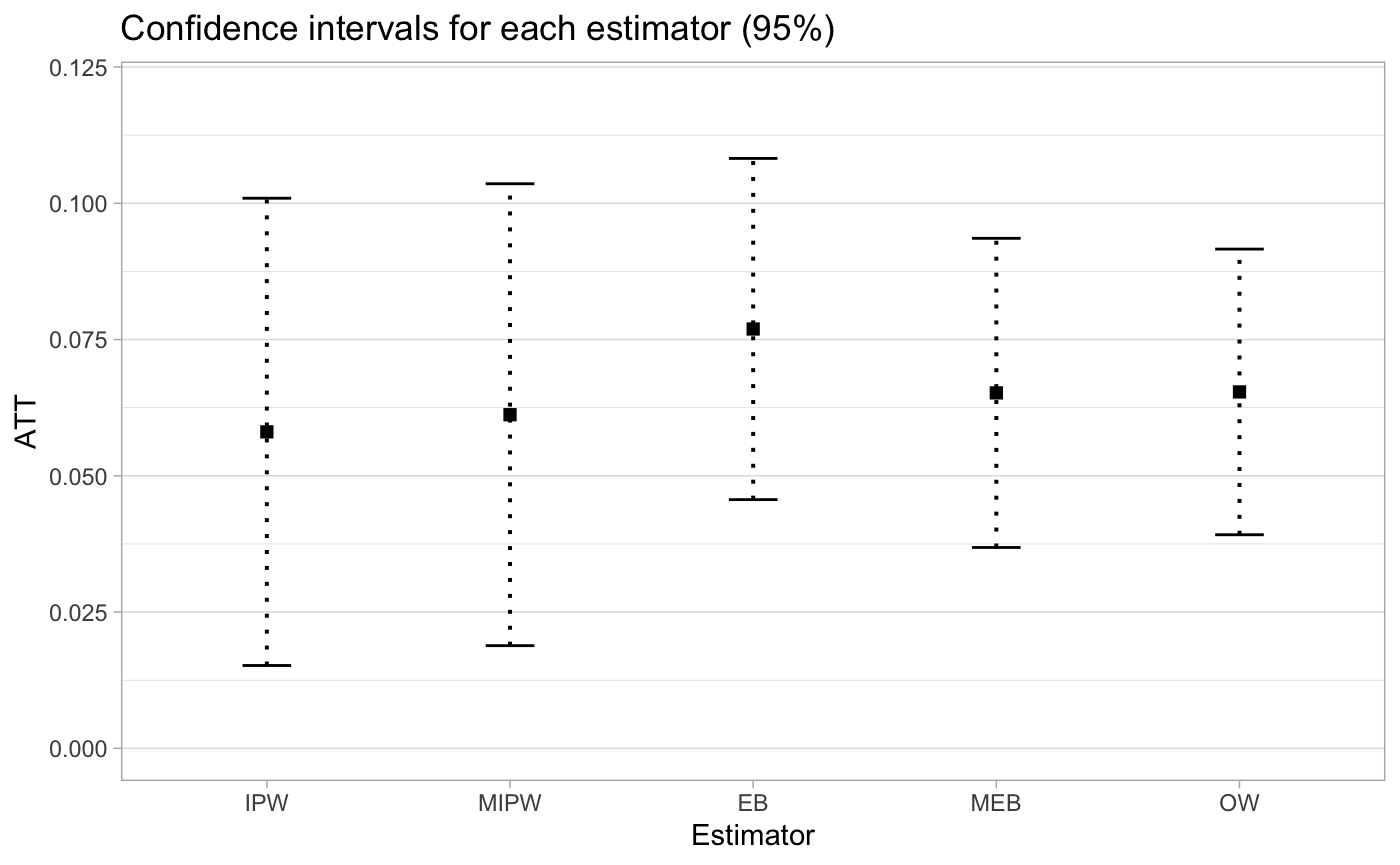}
     \caption{Confidence Interval constructed by the Bootstrap Standard Error.}\label{fig: 6}
   \end{minipage}
\end{figure}

From Figure \ref{fig: 5} and Table \ref{tab: 2}, we observe that, analogously, all of the estimators indicate that the RHC elevates the mortality rate of the patients regarding 95\% confidence interval from the standard error estimates. Even the overlap of the estimated propensity scores is sufficient, we can find, in Table \ref{tab: 2}, specific $\delta$ value for both IPW and EB such that standard error estimate reduces when mixing is applied. Regarding the guidelines of choosing the appropriate $\delta$ in Supplementary Material E, we take a closer examination in $\delta=0.1$ and $\delta=0.8$ for MIPW and MEB, respectively. Choosing the bootstrap standard error for its minimal condition requirements, the confidence interval are visualized in Fig. \ref{fig: 6}. We see that the similarity between MIPW and MEB are closer than IPW and EB. Moreover, MEB and OW are strikingly much alike to each other. Such observations strengthen the overall analyses from different methods on the same dataset. For a subsidiary note, when dealing with MIPW, just like any other estimators with large-sample property, it is up to the practitioners to decide a favorable standard error estimate considering the homogeneity assumption as one might prefer $\delta=0.7$ over $0.1$.

In addition, we also provided the degree of robustness in the estimated weights before and after applying mixing are illustrated as Figure 4 in Supplementary Material D. It shows that the mixing technique produces way more robust balancing weights in real-data. Integrating all the results, we conclude that mixing improves existing causal inference methodologies although the given sample is sufficient in overlap. We suggest the readers to apply mixing onto many techniques to tackle real datasets that suffer from the limited overlap.

\begin{table}[H]
    \centering
    \begin{tabular}{c|ccc}
        Estimator & Point Estimate & Robust SE & Boot SE \\
        \hline
        IPW & $0.05806$ & $0.02046$ & $0.02187$ \\
        MIPW$(0.1)$ & $0.06121$ & $0.01921$ & $0.02162$ \\
        MIPW$(0.2)$ & $0.06467$ & $0.01845$ & $0.02164$ \\
        MIPW$(0.3)$ & $0.06808$ & $0.01783$ & $0.02182$ \\
        MIPW$(0.4)$ & $0.07140$ & $0.01732$ & $0.02210$ \\
        MIPW$(0.5)$ & $0.07466$ & $0.01689$ & $0.02250$ \\
        MIPW$(0.6)$ & $0.07793$ & $0.01642$ & $0.02299$ \\
        MIPW$(0.7)$ & $0.08130$ & $0.01613$ & $0.02347$ \\
        MIPW$(0.8)^*$ & $0.08463$ & $0.01631$  & $0.02741$ \\
        MIPW$(0.9)$ & $0.08731$ & $0.07906$ & $0.02563$ \\
        EB          & $0.07693$ &           & $0.01597$  \\
        MEB$(0.1)$  & $0.07434$ &           & $0.01562$ \\
        MEB$(0.2)$  & $0.07262$ &           & $0.01533$ \\
        MEB$(0.3)$  & $0.07102$ &           & $0.01509$ \\
        MEB$(0.4)$  & $0.06958$ &           & $0.01486$ \\
        MEB$(0.5)$  & $0.06829$ &           & $0.01474$ \\
        MEB$(0.6)$  & $0.06720$ &           & $0.01462$ \\
        MEB$(0.7)$  & $0.06618$ &           & $0.01457$ \\
        MEB$(0.8)$  & $0.06521$ &           & $0.01447$ \\
        MEB$(0.9)$  & $0.06453$ &           & $0.01505$ \\
        OW          & $0.06539$ & $0.01327$ & $0.01337$  \\
        
    \end{tabular}
    \caption{ATT Estimate and Inference on the RHC Study. The asterisk $*$ means the estimator was calculated on approximate $\delta(=0.799)$ due to infeasible computed result on the exact $\delta(=0.8)$}
    \label{tab: 2}
\end{table}

\section{Discussion}
\label{sec6}

\subsection{Future Work}
In this paper, we have dealt with the simple mixed distribution and its application onto causal inference. The key feature of the simple mixed distribution is the shrinkage of the original propensity scores, toward the mean, $\pi$. However, as we have defined in Definition \ref{def 1}, the mixed distribution can leverage mixing proportion that can vary as a function.

Instead of ``pulling" the original propensity scores by a uniform constant ratio, one can aim to pull specific groups of units with insufficient common support to a greater extent. For instance, suppose we are given a sample with weak overlap, particularly when conditioned on a specific subgroup—for example, females. In this case, we apply a larger $\delta$ to mix the female units and a smaller $\delta$ to mix the male units, so that excessive information loss for weight shrinkage is avoided. We report the result of this toy example below; a more detailed description is provided in Supplementary Material F.

\begin{table}[H]
    \centering
    \begin{tabular}{c|cc}
    \Xhline{3\arrayrulewidth}
    Estimator & Bias & SD\\
    \hline
    \hline
    $IPW$ & $0.1812$ & $0.2656$\\
    Simple (Homogeneous) Mixing & $0.1590$ & $0.2440$\\
    Advanced (Heterogeneous) Mixing & $0.1497$ & $0.2232$\\
    \Xhline{3\arrayrulewidth}
    \end{tabular}
    \caption{Avenue for advanced mixing strategy}
    \label{tab: future work}
\end{table}

Table \ref{tab: future work} represents the finite-sample bias and standard deviation of the IPW applied with simple mixing strategy and the new mixing idea, that we will refer to as the ``advanced mixing". We observe that there exists advanced strategy that can outperform the simple mixing highlighting that there is still more to explore.

One potential limitation of the simple mixing strategy is that researchers have not yet established a theoretical guarantee of its efficiency gain when applied to IPW estimators. As future work, we aim to identify a more general mixing strategy—beyond the simple version—that can theoretically surpass standard causal estimators. Furthermore, we remain interested in exploring the application of mixing to a widely used approach in observational studies: \textit{matching}.

\subsection{Conclusion}
In this paper, we introduced mixing, a simple yet novel statistical tool designed to increase the overlap between the distributions of two groups, thereby facilitating covariate balancing. Through numerical studies, we demonstrated that applying this idea to causal inference methodology can help address one of the most fundamental assumption violations in observational studies—the overlap assumption—leading to more accurate inference.

Mixing is not merely a tool for handling weak overlap; it generally improves causal estimation without introducing the typical bias-variance trade-off, provided that the mixing proportion is appropriately selected. While the existing weighting literature proposes several alternatives with similar goals, these often require discarding samples or modifying the estimand—actions that may compromise external validity. In contrast, such sacrifices are not necessary when using mixing. Additionally, our method is flexible enough to be incorporated into a wide range of weighting estimators.

Given its many advantages, we encourage practitioners to apply mixing to various causal estimators, particularly when weak overlap is present. Furthermore, its elementary mathematical structure leaves ample room for future improvement and theoretical development.

\bibliographystyle{apalike}

\bibliography{references}
\end{document}